\documentclass[twocolumn]{aastex631}
\usepackage{amsmath}
\usepackage{wrapfig}
\usepackage{mathtools}

\begin{document}

\title{Formation of Merging Stellar-Mass Black Hole Binaries by Gravitational Wave Emission in Active Galactic Nucleus Disks}

\correspondingauthor{Barak Rom}
\email{barak.rom@mail.huji.ac.il}

\author[0000-0002-7420-3578]{Barak Rom}
\affiliation{Racah Institute of Physics, The Hebrew University of Jerusalem, 9190401, Israel}
\author[0000-0002-1084-3656]{Re'em Sari}
\affiliation{Racah Institute of Physics, The Hebrew University of Jerusalem, 9190401, Israel}
\affiliation{Research Center for the Early Universe, Graduate School of Science, University of Tokyo, Bunkyo-ku, Tokyo 113-0033, Japan}
\author[0000-0002-1934-6250]{Dong Lai}
\affiliation{Center for Astrophysics and Planetary Science, Department of Astronomy, Cornell University, Ithaca, NY 14853, USA}
\affiliation{Tsung-Dao Lee Institute, Shanghai Jiao-Tong University, Shanghai, 520 Shengrong Road, 201210, China
}

\shorttitle{Binary capture by GW emission}
\shortauthors{Rom, Sari \& Lai}

\begin{abstract}
\noindent 
Many stellar-mass Black Holes (sBHs) are expected to orbit supermassive black holes at galactic centers. For galaxies with Active Galactic Nuclei (AGN), it is likely that the sBHs reside in a disk.
We study the formation of sBH binaries via gravitational wave emission in such disks. 
We examine analytically the dynamics of two sBHs orbiting a supermassive black hole, estimate the capture cross section, and derive the eccentricity distribution of bound binaries at different frequency bands.
We find that the majority of the merging sBH binaries, assembled in this manner, can be measured as highly eccentric, detectable in the LIGO-Virgo-KAGRA (LVK) band from their formation, with $(1-e)\ll1$, through their circularization and up to their merger; the remaining binaries circularize to small eccentricities ($e\lesssim0.3$) before entering the LVK band.
More eccentric mergers would be observed for sBHs with higher random velocities, closer to the supermassive black hole, or at lower observing frequency bands, as planned in future gravitational wave detectors such as the Einstein Telescope and LISA.
\end{abstract}
 
\keywords{Gravitational wave sources (677), Active galactic nuclei (16), Stellar mass black holes (1611), Supermassive black holes (1663)}

\section{Introduction}

The detections of Gravitational Waves (GWs) from merging binary Black Holes (BHs) by the LIGO-Virgo-KAGRA (LVK) collaboration \citep{LIGO1, LIGO_3} open a new era in astronomy. 
One of the main questions arising from these first detections concerns the origins of the observed binaries and the mechanisms that drive them to merge within a Hubble time. 

Several formation channels have been proposed, from isolated binary evolution
\citep[e.g.,][]{Lipunov1997,Podsiadlowski2003,
Mandel_2016,Belczynski_2016},
to tertiary-induced mergers (via Lidov-Kozai effect) in stellar triple/quadrupole systems
\citep[e.g.,][]{Silsbee2017,Liu2018,Liu2019,
Fragione2019}
or in binaries around central supermassive BHs
\citep[e.g.,][]{Antonini2012, Petrovich2017, Hoang_2018, Liu2019b, Liu_2021},
and dynamical captures in dense stellar environments, invoking, for example, multiple stellar systems dynamics \citep[e.g.,][]{Samsing_2014,Rodriguez2015, Antonini_2016} and GW emission during close encounters \citep[e.g.,][]{OLeary_2009,Gondan_2018,Samsing_2020}.

Another promising formation channel considers Active Galactic Nuclei (AGN) as a fertile ground for binary BH mergers
\citep[e.g.,][]{McKernan2012, Stone_2016, Bartos_2017, Tagawa_2020b, Ford_2022}.
Binary stellar-mass BHs (sBHs) in flat disks may be hardened, or even driven to merger, by a series of nearly coplanar binary-single scatterings \citep[e.g.,][]{Stone_2016, Leigh2018, Samsing_2022}. Hydrodynamical interaction between the gaseous AGN disk and an embedded BH binary may lead to orbital contraction of the binary under a variety of conditions
\citep[][]{LiYP2021ApJ,LiYP2022ApJL,Dempsey-Li-2022,Li-Lai-2022,Li-Lai-2023}.
Binary BHs may form from singles due to GW emission in very close encounters \citep{Li_Lai_Rodet_2022,Boekholt_2023}
or due to gas drags
\citep{Tagawa_2020b,Li-Dempsey-2023, DeLaurentiis_2023,Rozner_2023,Qian_2023}. 

In this work, we study analytically the characteristics of sBH binary capture via GW emission. 
Considering the expected settings in AGN disks, the interplay between dissipation induced by the gaseous disk \citep{Syer_1991,Generozov_2023} and excitation due to close encounters between BHs leads schematically to two possible steady-state distributions of the embedded sBHs: one with small eccentricities and negligible inclinations, where the dynamics are shear-dominated, while the other includes significant eccentricities and inclinations, and the dynamics are dispersion-dominated \citep{Goldreich_04,Trani_2023}.
In the first case, the relative velocities between the sBHs are determined by the Keplerian shear between circular orbits at different radii, while in the second case, the relative velocities stem from the velocity dispersion.

Following \cite{Li_Lai_Rodet_2022}, we study the close encounter dynamics of two sBHs orbiting a Super-Massive Black Hole (SMBH), without taking into account the effects of the surrounding gas nor accretion into the sBHs.
The gas effects are complex, and generally require hydrodynamical simulations for full 
treatment \citep{Li-Dempsey-2023,Rowan_2022}.

In Section \ref{sec:3b}, we formulate the restricted three-body problem with an effective GW-induced friction force. In Section \ref{sec:tdgw}, we analyze analytically the orbital evolution given a shear-dominated dynamics. In Sections \ref{sec:e} and \ref{sec:e2}, we derive the bound binaries' eccentricity distribution as a function of the semimajor axis for shear-dominated and dispersion-dominated dynamics, respectively. Finally, we summarize our conclusions in section \ref{sec:sum}.

\section{The circular restricted three-body problem} \label{sec:3b}

We study the shear-dominated dynamics of two sBHs, with masses $m_{1,2}$, orbiting around a SMBH, with mass $M$, along nearly coplanar circular orbits, with radii $r_{1,2}$. We denote the binary total mass, $m=m_1+m_2\ll M$, and the center-of-mass radius $a=\left(m_1r_1+m_2r_2\right)/m$. 

In the frame rotating at $\Omega=\left(GM/a^3\right)^{1/2}$, the sBHs relative position, $\vec{r}=\vec{r}_1-\vec{r}_2$, is governed by the well-known Hill equations\footnote{Although the Hill equations were originally derived assuming a mass hierarchy, $m_2\ll m_1\ll M$, they are applicable for more general cases, e.g., for arbitrary mass ratio between $m_1$ and $m_2$, as long as $m_1+m_2\ll M$ \citep[for more details see][]{Henon_1986}.} \citep{Hill_1878,Henon_1986,Murray_Dermott_2000,Sari_2009}
\begin{equation} \label{eq:tb2}
    \left\{\begin{aligned}
        &\ddot{x}=2\dot{y}+3x\left(1-\frac{1}{r^3}\right)-f_x, \\
        & \ddot{y}=-2\dot{x}-\frac{3y}{r^3}-f_y,
    \end{aligned}\right.
\end{equation}
where $\vec{f}=\left(f_x,f_y\right)$ is any additional dissipative force, as discussed below. 
Note that in Eq. (\ref{eq:tb2}), the $\hat{x}$-axis points from the SMBH to the center of mass of the sBHs. 
The coordinates are in units of the Hill radius, $R_H=a\left[m/\left(3M\right)\right]^{1/3}$, and time is in units of $\Omega^{-1}$, so the velocities are in units of $v_H=\Omega R_H=\sqrt{Gm/\left(3R_H\right)}$.

We define the impact parameter, $b$, as the initial radial offset between the two sBHs and solve Eq. (\ref{eq:tb2}) with the following initial conditions: $x\left(t=0\right)=b$, $y\left(t=0\right)\gg b$, and $\vec{v}\left(t=0\right)=-\left(3b/2\right)\hat{y}$.

\subsection{Gravitational-wave``friction" force} 
The GW emission leads to a dissipation of energy and angular momentum. We model this effect by introducing in Eq. (\ref{eq:tb2}) an effective GW-induced ``friction" force of the form
\begin{equation} \label{eq:gw1}
    \frac{\vec{f}}{\mu v_H^2/R_H}=\kappa\left(\frac{r}{R_H}\right)^{-9/2}\hat{v},
\end{equation}
where $\mu=m_1m_2/m$ is the reduced mass, and $\hat{v}$ is the unit vector.
We determine $\kappa$ by demanding that the energy loss due to the friction force reproduces the correct orbital-averaged GW energy loss along a highly eccentric orbit \citep{Peters_64}
\begin{equation} \label{eq:gw2}
    \frac{\Delta E_{GW}}{\mu v_H^2}=\frac{85\pi}{32}\left(\frac{\mu}{m}\right)\left(\frac{R_s}{R_H}\right)^{5/2}\left(\frac{r_p}{R_H}\right)^{-7/2},
\end{equation}
where $r_p$ is the periapsis and $R_s=2Gm/c^2$.

Comparing Eq. (\ref{eq:gw2}) with the work done by the friction force, Eq. (\ref{eq:gw1}), along a parabolic orbit yields
\begin{equation} \label{eq:gw3}
    \kappa=\frac{2975\pi}{2048}\left(\frac{\mu}{m}\right)\left(\frac{R_s}{R_H}\right)^{5/2}.
\end{equation}
The value of $\kappa$ span over several order of magnitudes, as it depends on the binary's symmetric mass ratio, $\mu/m$, the mass ratio of the binary and the SMBH, $m/M$, and the binary's center-of-mass orbital radius, $a$, such that
\begin{equation}
    \kappa\simeq4\times10^{-16}\left(\frac{10^3}{\widetilde{a}}\right)^{5/2}\left(\frac{m/M}{10^{-5}}\right)^{5/3}\left(\frac{\mu/m}{1/4}\right),
\end{equation}
where $\tilde{a}=a/\left(2GM/c^2\right)$, is the binary's orbital radius in units of the SMBH's Schwarzschild radius.

Note that although this simplified description of the friction force is not valid for general orbits, it faithfully describes the GW emission effect on highly eccentric orbits\footnote{This formalism is valid for circular orbits as well, for which the numerical coefficient in Eq. (\ref{eq:gw3}) is $12\sqrt{2}/5$. However, when considering the full eccentricity evolution, this simplified description deviates by order unity from the exact result \citep{Peters_64}.}, where the GW emission is most efficient during the pericenter passage. Thus, this prescription is adequate for the study of sBHs captures during close encounters, as the sBHs initially approach each other along roughly parabolic orbits. 

A bound binary can be formed by emitting GWs, during pericenter passage, that lead to an energy loss comparable to the Hill energy \citep{Samsing_2018,Tagawa_2021,Li_Lai_Rodet_2022,Boekholt_2023}, $\Delta E_{GW}\sim E_H= \mu v_H^2$. 
Therefore, the critical periapsis distance between the two sBHs required for binary formation is
\begin{equation}\label{eq:gw5}
    \frac{r_{p,\rm cap}}{R_H}\sim\kappa^{2/7}.
\end{equation}

The majority of the captured orbits lies in a narrow band of width $\Delta b_c$, around the ``zero-angular momentum" impact parameter, $b_0$, which corresponds to a direct plunge trajectory. Deep enough in the Hill sphere, the tides from the SMBH are negligible so the binary angular momentum is practically conserved. For the ``zero-angular momentum" impact parameter, the binary's angular momentum vanishes and the two sBHs arrive arbitrarily close.

Consider $m_1$ and $m_2$ approaching each other at $r\sim R_H$ with a small impact parameter $\left|\Delta b\right|\leq\Delta b_c$, such that they reach periastron separation $r_p\ll R_H$. Angular momentum conservation gives $v_H \left|\Delta b\right|\simeq \sqrt{2Gmr_p}$, which implies
\begin{equation} \label{eq:gw6}
    \frac{r_p}{R_H}\sim\left(\frac{\Delta b}{R_H}\right)^2.
\end{equation}

\begin{figure}[ht!]
    \centering   
    \includegraphics[width=8.6cm]{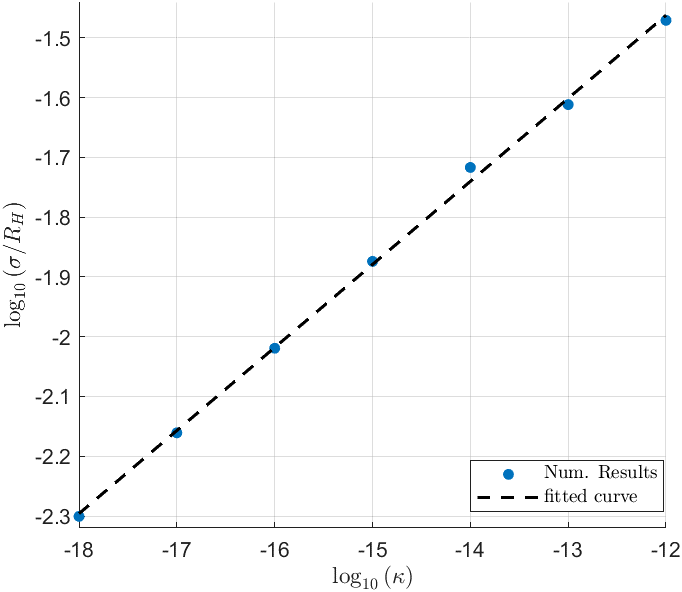}    
    \caption{A power-law fit for the numerically calculated capture cross-section as a function of the friction parameter $\kappa$ (defined in Eq. \ref{eq:gw1}). The fit yields an exponent of $0.14\pm0.01$, consistent with the analytical value of $1/7$, given in Eq. (\ref{eq:gw9}).}
    \label{fig:1a}
\end{figure}

In this band of impact parameters, assuming that the orbits of the two sBHs are perfectly aligned, the probability density function (pdf) for capture satisfies $p(\Delta b)=1/\Delta b_c$. Using Eq. (\ref{eq:gw6}), we get that the pdf with respect to the periapsis is
\begin{equation}\label{eq:gw7}
    p(r_p)\propto\left(\frac{r_p}{r_{p,\rm cap}}\right)^{-1/2},
\end{equation}
and the cumulative distribution function (cdf), namely, the probability for capture with periapsis smaller than $r_p$, is
\begin{equation}\label{eq:gw8}
    P(<r_p)=\left(\frac{r_p}{r_{p,\rm cap}}\right)^{1/2}\simeq\kappa^{-1/7}\left(\frac{r_p}{R_H}\right)^{1/2},
\end{equation}
in agreement with previous results \citep{Li_Lai_Rodet_2022,Boekholt_2023}.
We define the one-dimensional cross section $\sigma$ as the linear measure of the impact parameters that lead to capture. Therefore, from Eqs. (\ref{eq:gw5}) and (\ref{eq:gw6}), the cross-section for capture is given by
\begin{equation}\label{eq:gw9}
    \frac{\sigma}{R_H}=2\sqrt{\frac{r_{p,\rm cap}}{R_H}}=1.8\kappa^{1/7}.
\end{equation}
The order unity coefficient in Eq. (\ref{eq:gw9}) is determined numerically, as discussed below (and see Fig. \ref{fig:1a}).

\begin{figure}[ht!]
    \centering
    \includegraphics[width=8.6cm]{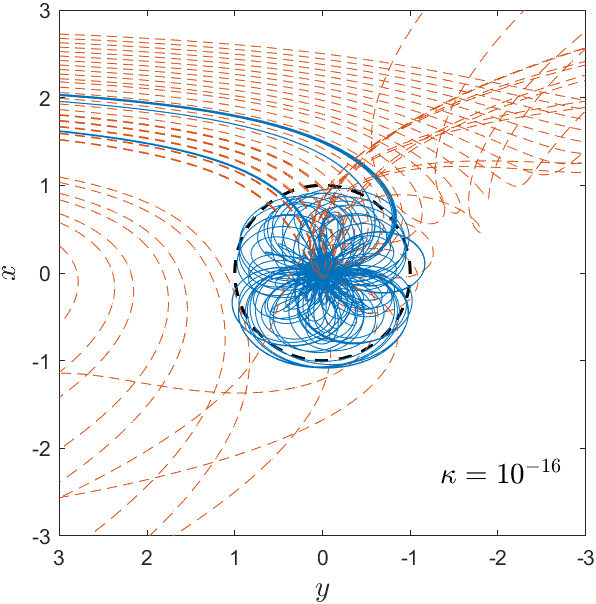}    
    \caption{The circular restricted three-body problem with GW emission: sample trajectories for different impact parameters in the corotating frame. The blue lines are the captured orbits, while the dashed red lines present a part of the unbound trajectories. We assume that initially the two sBHs are on coplanar, circular orbits around the SMBH. The Hill sphere is marked by a black dashed line, and the friction parameter is $\kappa=10^{-16}$ (see Eq. \ref{eq:gw1}).}
    \label{fig:1}
\end{figure}

Following \citet{Goldreich_2002}, we validate our prediction by numerically solving Eq. (\ref{eq:tb2}), with the effective GW friction force, Eq. (\ref{eq:gw1}), for different impact parameters. Figure \ref{fig:1} presents an example of the orbits obtained for $\kappa=10^{-16}$. We repeat this calculation for different values of $\kappa$ and estimate the capture cross-section, as defined above. We get the expected power-law trend, as presented in Fig. \ref{fig:1a}, with an exponent of $0.14\pm0.01$, consistent with the exponent of $1/7$ in Eq. (\ref{eq:gw9}). Moreover, we extract from the numerical calculation the order unity coefficient, as appear in the second equality of Eq. (\ref{eq:gw9}). Furthermore, we find that the majority of the captured orbits are concentrated in two narrow bands of impact parameters, around $b_0=2.08$ and $b_0=2.39$, that correspond to the ``zero-angular momentum" trajectories as expected.  

In addition to the abovementioned captured trajectories, which become bound already after the first pericenter passage, there are long-lived orbits that result in a capture, i.e., trajectories where the two sBHs stay inside their mutual Hill sphere for several orbits before forming a bound binary \citep[as suggested by][in the context of Kuiper Belt binaries]{Astakhov_2005,Lee_2007}. However, since these trajectories are exponentially rare \citep{Schlichting_2008, Boekholt_2023}, their contribution to the cross-section is subdominant.

Our calculation assumes that a capture requires an energy loss of $E_H=\mu v_H^2$. In contrast, \cite{Goldreich_2002} showed that even for smaller loss of energy there are ``almost-bound" orbits that can be captured. These provide the majority of captures in their Kuiper belt binary model where the dynamical friction is weak. The reason that such ``almost-bound" orbits are responsible for the majority of captures in the dynamical friction case but are not important in the GW capture case is that these orbits are not centered around the zero-angular momentum orbit, namely, they do not have small periapsis, compared to the Hill radius, and so the amount of GWs they emit is negligible and they remain unbound. 
\section{Tidal forces versus Gravitational Waves} \label{sec:tdgw}

\begin{figure}[ht!]
    \centering
    \includegraphics[width=8.6 cm]{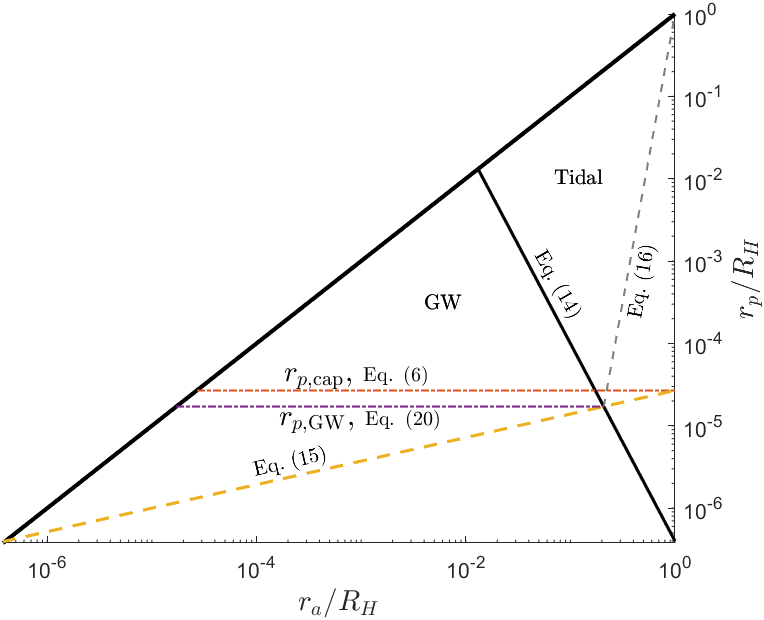}
    \caption{Orbital evolution regions in the $\left(r_a,r_p\right)$ phase space, where $r_a$ is the semimajor axis and $r_p$ is the periapsis of the relative orbit of the sBHs, for a given friction parameter $\kappa=10^{-16}$ (see Eq. \ref{eq:gw1}). The diagonal black line distinguishes between the tidal-force-dominated region to the GW-dominated one, Eq. (\ref{eq:tg4}). The horizontal red dashed-dotted line represents the maximal periapsis needed for capture, Eq. (\ref{eq:gw5}); the horizontal purple dashed-dotted line represents the maximal periapsis from which the eccentricity evolves due to GW emission, Eq. (\ref{eq:e3a}); the yellow dashed line traces the evolution of initially parabolic orbits after their first pericenter passage, Eq. (\ref{eq:tg5}); the gray dashed line marks the boundary of the possible plunge region, Eq. (\ref{eq:tg6}).}
    \label{fig:2}
\end{figure}

The orbital evolution of the sBHs is driven by the tidal force, exerted by the SMBH, and the GW emission associated with their relative motion.
Considering orbits that enter the Hill sphere, we can define their semimajor axis $r_a\lesssim R_H$, periapsis $r_p$, and angular momentum $J(r_p)\sim \mu\sqrt{Gmr_p}$. At large separations the tidal force dominates, leading to an effective diffusion in angular momentum, while at small separations the GW emission prevails, leading to a fast circularization. The division of the phase space, according to these different regions, is depicted in Fig. \ref{fig:2}.

In one orbital period, $T(r_a)\sim\sqrt{r_a^3/\left(Gm\right)}$, the tidal force changes the orbital angular momentum by\footnote{Depending on the orientation of the sBHs relative to the SMBH, the numerical prefactor in Eq. (\ref{eq:tg1}) may vary from 0 up to $\sim30$.}
\begin{equation}\label{eq:tg1}
    \Delta J(r_a) \sim \frac{GM\mu r_a^2}{a^3}T(r_a)\sim\mu v_H R_H\left(\frac{r_a}{R_H}\right)^{7/2},
\end{equation}
in a random direction. Therefore, the characteristic timescale for an order unity change of the angular momentum is
\begin{equation} \label{eq:tg2}
    \tau_{\rm tidal}\sim T(r_a)\left[\frac{J(r_p)}{\Delta J(r_a)}\right]^2\sim\frac{R_H}{v_H}\frac{r_p}{R_H}\left(\frac{r_a}{R_H}\right)^{-11/2}.
\end{equation}

On the other hand, the GW timescale, for changing the orbital energy, is \citep{Peters_64}
\begin{equation}\label{eq:tg3}
\begin{aligned}
    \tau_{GW}&\sim T(r_a)\frac{m}{\mu}\frac{r_p^{7/2}}{R_s^{5/2}r_a}\\
    &\sim\frac{R_H}{v_H}\frac{m}{\mu}\left(\frac{R_s}{R_H}\right)^{-5/2}\left(\frac{r_p}{R_H}\right)^{7/2}\left(\frac{r_a}{R_H}\right)^{1/2}.
\end{aligned}
\end{equation}

\begin{figure}[ht!]
    \centering
    \includegraphics[width=8.6 cm]{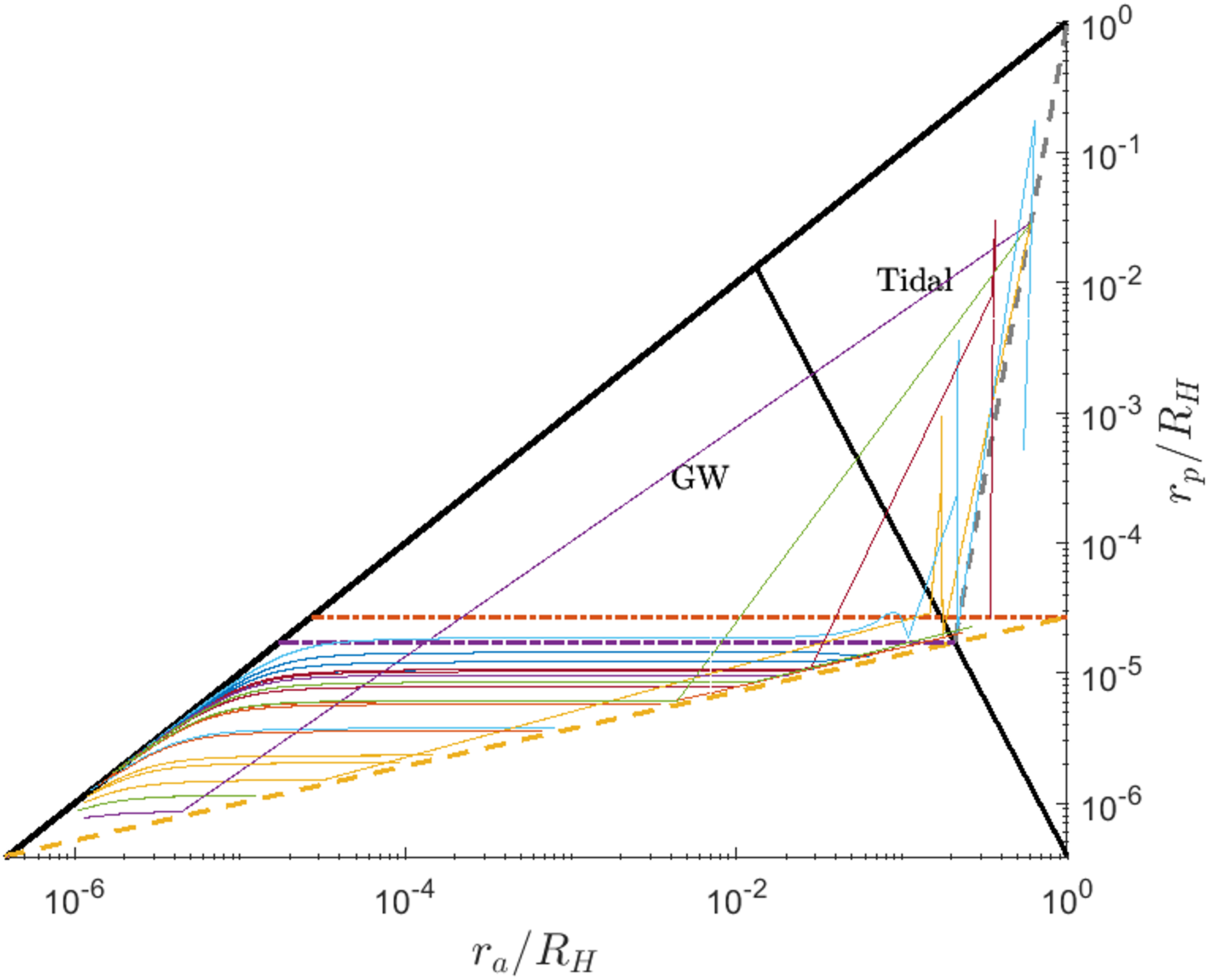}
    \caption{A sample of numerically calculated captured orbits, presented in the $(r_a,r_p)$ plane. The orbits were calculated by solving Hill equations, Eq. (\ref{eq:tb2}), with GW friction force, Eq. (\ref{eq:gw1}), and are presented by the colored lines.
    As expected, in the GW-dominated region, below the diagonal black line, the orbit undergo circularization at approximately constant periapsis, while in the tidal-force-dominated region, above the diagonal line, the orbit's periapsis changes stochastically at a roughly constant semimajor axis. As in Fig. \ref{fig:2}, the horizontal red dashed-dotted line represents the maximal periapsis needed for capture, Eq. (\ref{eq:gw5}) and the yellow dashed line traces the evolution of initially parabolic orbits after their first pericenter passage, Eq. (\ref{eq:tg5}).}
    \label{fig:3}
\end{figure}

Equating the two timescales gives
\begin{equation}\label{eq:tg4}
       \frac{r_p}{R_H}\sim\kappa^{2/5}\left(\frac{r_a}{R_H}\right)^{-12/5},
\end{equation}
which behaves as an effective separatrix in the $(r_a,r_p)$ space: above it, the tidal force timescale is shorter, leading to a random walk in the $r_p$ direction, with a roughly constant $r_a$; below it, the GW emission dominates the evolution, causing a rapid decrease of $r_a$ with approximately constant $r_p$. We verify this analytical prediction with numerically calculated orbits and present in Fig. \ref{fig:3} a sample of the captured orbits. 

We can further delimit the phase space by noting that the relevant captured orbits initially enter the Hill sphere along an approximately parabolic orbit. Therefore, after the first pericenter passage, their new semimajor axis is determined by the requirement $\Delta E_{GW}(r_p)\sim\left|E(r_a)\right|\sim Gm\mu/r_a$. Therefore, they settle to an orbit along a line which is given by
\begin{equation} \label{eq:tg5}
        \frac{r_p}{R_H}\sim\left(\kappa\frac{r_a}{R_H}\right)^{2/7}.
\end{equation}

Finally, we note that a direct plunge is possible when the change in angular momentum due to the tidal force in one orbit is comparable to the orbital angular momentum, i.e., 
$\Delta J(r_a)\sim J(r_p)$, or
\begin{equation} \label{eq:tg6}
   \frac{r_p}{R_H}\sim \left(\frac{r_a}{R_H}\right)^7.
\end{equation}

\section{Eccentricity Distribution} \label{sec:e}

We calculate the eccentricity distribution at a given binary semimajor axis, $r_a$, and estimate the probability to retain a non-negligible eccentricity in the LVK band, $\sim10-1000\rm Hz$ \citep{Virgo,Martynov_2016,LIGO_2018,KAGRA}.

At small eccentricities, the GW emission is nearly monochromatic, with the GW frequency twice the orbital frequency, $f=\pi^{-1}\left(Gm/r_a^3\right)^{1/2}$, or equivalently
\begin{equation}\label{eq:e1}
    \frac{r_{a}(f)}{R_s}\approx 13 \left(\frac{10\rm Hz}{f}\right)^{2/3} \left(\frac{50M_\odot}{m}\right)^{2/3}.
\end{equation}
On the other hand, the GW emission from eccentric orbits is broadband \citep{Peters_1963,Turner_1977}, as further discussed in Section \ref{sec:strain}.

The eccentricity evolution from capture, at the initial periapsis $r_{p,i}$ and eccentricity $e_i\approx1$, up to the relevant semimajor axis, $r_{a}(f)$, can be evaluated analytically using \cite{Peters_64} result \citep{Samsing_2018,Linial_2023}:
\begin{equation}\label{eq:e2}
    \frac{r_{a}(f)}{r_{p,i}}=g(e),
\end{equation}
where
\begin{equation}\label{eq:e3}
    g(e)=\frac{2e^{12/19}}{1-e^2}\left(\frac{304+121e^2}{425}\right)^{870/2299}.
\end{equation}

The minimal eccentricity at a given frequency, $e_{\rm min}$, arises from orbits that were captured at the maximal initial periapsis in the GW-dominated region,
\begin{equation} \label{eq:e3a}
    \begin{aligned}
    \frac{r_{p,\rm GW}}{R_H}\sim\kappa^{14/47},   
    \end{aligned}
\end{equation}
which introduces a characteristic frequency
\begin{equation}\label{eq:e6}
    \begin{aligned}
        f_{c}&=\frac{1}{\pi}\sqrt{\frac{Gm}{r_{p,\rm GW}^3}}\approx 2{\rm Hz} \left(\frac{10^3}{\widetilde{a}}\right)^{18/47}\\
        &\times  \left(\frac{50M_\odot}{m}\right)\left(\frac{m/M}{10^{-5}}\right)^{12/47}\left(\frac{1/4}{\mu/m}\right)^{21/47}.
    \end{aligned}
\end{equation}
Note that $r_{p,\rm GW}$ is slightly smaller than $r_{p,\rm cap}$, Eq. (\ref{eq:gw5}). The value of $r_{p,\rm cap}$ is set by Eq. (\ref{eq:tg5}) with $r_a\sim R_H$, while $r_{p,\rm GW}$ is the intersection of Eqs. (\ref{eq:tg4}) and (\ref{eq:tg5}). 

Depending on the frequency, we obtain two qualitatively different possibilities for the eccentricity distribution.
In the first case, all of the binaries retain a significant eccentricity. This occurs if $r_{a}(f)\gtrsim r_{p,\rm GW}$, or equivalently $f\lesssim f_{c}$, and thus, using Eq. (\ref{eq:e2}), the minimal eccentricity satisfies $e_{\rm min}\gtrsim0.33$, since  
\begin{equation}\label{eq:emin}
    g(e_{\rm min})=\left(\frac{f_c}{f}\right)^{2/3},
\end{equation}
as follows from the definition of $e_{\rm min}$ together with Eqs. (\ref{eq:e2}) and (\ref{eq:e6}).

This possibility can be valid at higher frequencies if we consider a less massive SMBH, or a sBH binary with smaller mass ratio, total mass, or orbital radius, as evident from Eq. (\ref{eq:e6}). 
The corresponding minimal eccentricity satisfies
\begin{equation}\label{eq:e6a}
    \begin{aligned}
        \left(1-e_{\rm min}\right)\approx& 0.1 \left(\frac{f}{0.05\rm Hz}\right)^{2/3}\left(\frac{\widetilde{a}}{10^3}\right)^{12/47}\left(\frac{m}{50M_\odot}\right)^{2/3}\\
        & \times \left(\frac{10^{-5}}{m/M}\right)^{8/47}\left(\frac{\mu/m}{1/4}\right)^{14/47}.
    \end{aligned}
\end{equation}

In the second case, the majority of the orbits have small eccentricities, at the given orbital frequency. 
Namely, in this case $r_a(f)\ll r_{p,\rm GW}$ (i.e., $f\gg f_c$) and accordingly $e_{\rm min}\ll1$,
\begin{equation}\label{eq:e4}
    \begin{aligned}
        e_{\rm min}\approx& 0.01 \left(\frac{50\rm Hz}{f}\right)^{19/18}\left(\frac{10^3}{\widetilde{a}}\right)^{19/47}\left(\frac{50M_\odot}{m}\right)^{19/18}\\
        & \times \left(\frac{m/M}{10^{-5}}\right)^{38/141}\left(\frac{1/4}{\mu/m}\right)^{133/282}.
    \end{aligned}
\end{equation}

We note that at a given semimajor axis, $r_a(f)$, there is a maximal eccentricity, $e_{\rm max}$, determined by the initial periapsis, $r_p(f)$, as given by Eq. (\ref{eq:tg5}). Namely, after one pericenter passage, the roughly parabolic orbit settles to a bound orbit with $r_a(f)$ as its semimajor axis. Therefore, $e_{\rm max}$ satisfies
\begin{equation}\label{eq:emax}
    \begin{aligned}
        \left(1-e_{\rm max}\right)\approx& 0.2\left(\frac{f}{10 \rm Hz}\right)^{10/21}\left(\frac{m}{50M_\odot}\right)^{10/21}\left(\frac{\mu/m}{1/4}\right)^{2/7}.
    \end{aligned}
\end{equation}
A smaller initial periapsis leads to the formation of a binary with smaller semimajor axis, i.e., higher frequency.

Considering the GW dominated region, the pericenter distribution follows the same power law as Eq. (\ref{eq:gw8}), but normalized by $r_{p,\rm GW}$ rather than $r_{p,\rm cap}$. Therefore, at a given orbital frequency, the eccentricity cdf, using Eq. (\ref{eq:e2}), is given by
\begin{equation}\label{eq:Pe}
    \begin{aligned}
        P(\geq e)\simeq &\left(\frac{f_c}{f}\right)^{1/3}\sqrt{\frac{1}{g(e)}}\\
        &\times\frac{\sqrt{g\left(e\right)}-\sqrt{g\left(e_{\rm max}\right)}}{\sqrt{g\left(e_{\rm min}\right)}-\sqrt{g\left(e_{\rm max}\right)}},
    \end{aligned}
\end{equation}
where $e_{\rm min}$ and $e_{\rm max}$ are given in Eqs. (\ref{eq:emin}) and (\ref{eq:emax}).
\begin{figure}[ht!]
    \centering
    \includegraphics[width=8.6 cm]{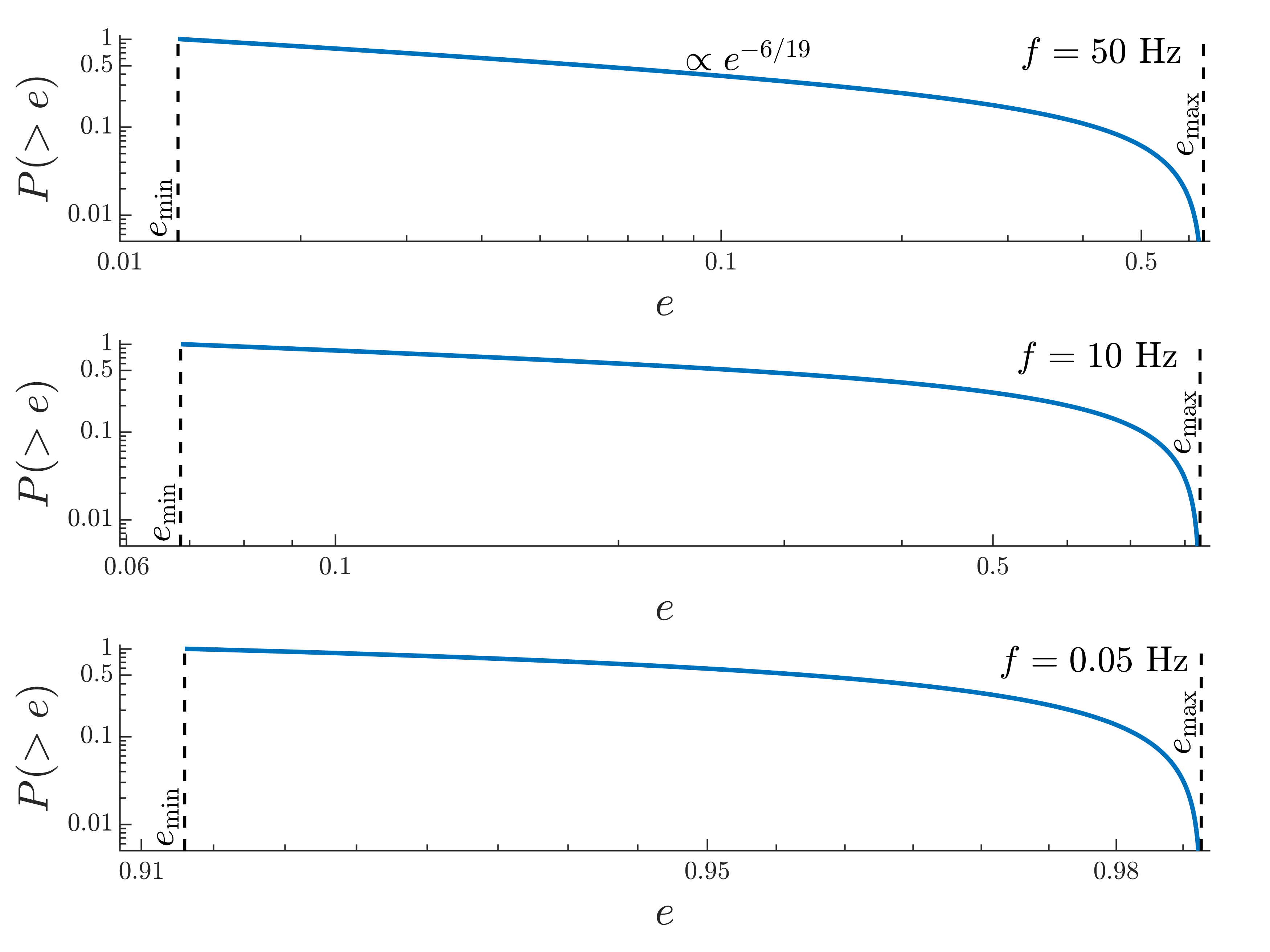}
    \caption{The cumulative distribution function at different frequencies: $50\rm Hz$ (top panel); $10\rm Hz$ (middle panel); and $0.05\rm Hz$ (lower panel), as given by Eq. (\ref{eq:Pe}). The lower and top panels exhibit edge cases where the minimal eccentricity is significant (Eq. \ref{eq:e6a}) or negligible (Eq. \ref{eq:e4}), respectively. In this calculation we assume an equal-mass binary with total mass $m=50M_\odot$, orbiting an SMBH with $M=5\cdot10^6M_\odot$, on a circular orbit at $10^3$ of the SMBH's Schwarzschild radius.}
    \label{fig:5}
\end{figure}

We present in Fig. \ref{fig:5} the eccentricity distribution at different frequencies, given our fiducial parameters.
We note that for $e\ll1$, Eq. (\ref{eq:Pe}) yields 
\begin{equation}\label{eq:e8}
    \begin{aligned}
        P(\geq e)\simeq& \left(\frac{f_c}{f}\right)^{1/3}\sqrt{\frac{1}{g\left(e\right)}}\\
        \approx&\frac{0.3}{e^{6/19}}\left(\frac{50\rm Hz}{f}\right)^{1/3}\left(\frac{10^3}{\widetilde{a}}\right)^{6/47}\left(\frac{50M_\odot}{m}\right)^{1/3}\\
        & \times \left(\frac{m/M}{10^{-5}}\right)^{4/47}\left(\frac{1/4}{\mu/m}\right)^{7/47},
    \end{aligned}
\end{equation}
On the other hand, for $e\simeq e_{\rm max}\simeq1$
\begin{equation}\label{eq:e8_1}
    \begin{aligned}
        P(\gtrsim e\simeq1)&\simeq\left(\frac{f_c}{f}\right)^{1/3}\sqrt{1-e_{\rm max}}\left[\sqrt{\frac{1-e}{1-e_{\rm max}}}-1\right].
    \end{aligned}
\end{equation}

\begin{figure*}[ht!]
    \centering
    \includegraphics[width=\textwidth]{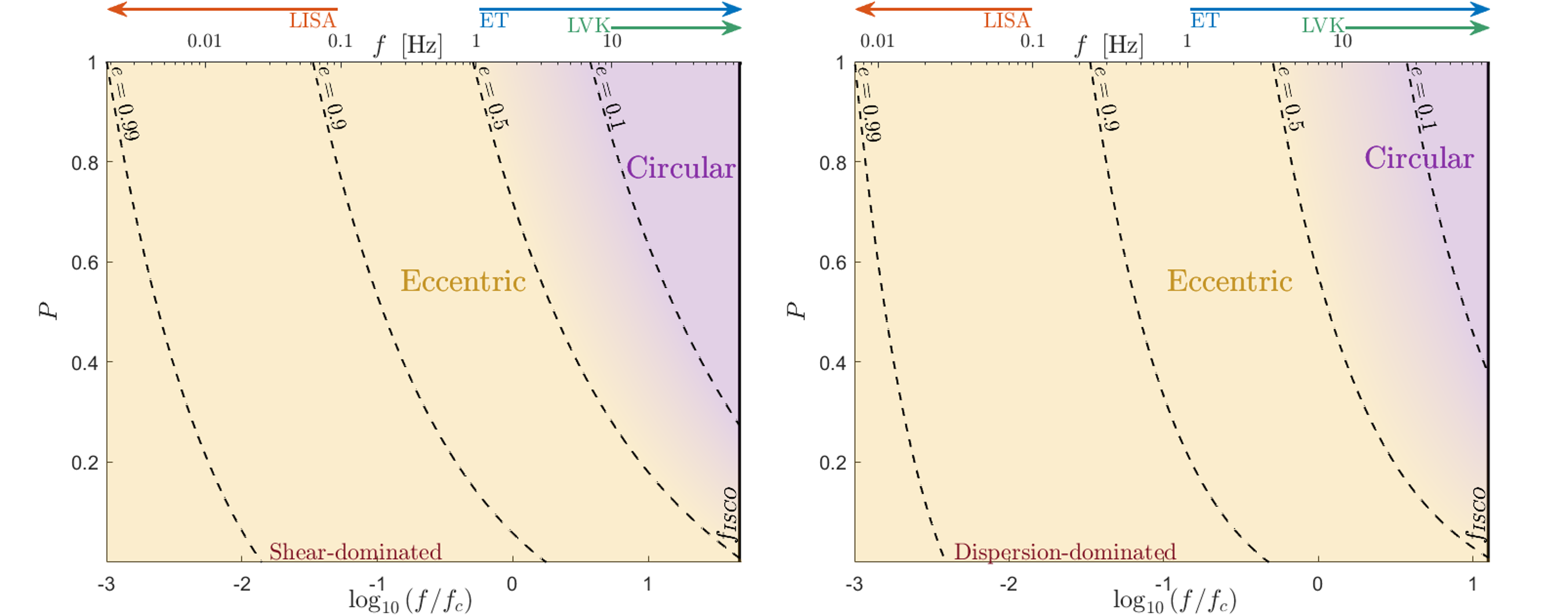}
    \caption{The eccentricity probability distribution of sBH binaries as a function of the orbital frequency, assuming shear-dominated (left panel) or dispersion-dominated (right panel) dynamics. In a given frequency, the colored regions represent the probability to observe circular orbits (purple region), or eccentric orbits (yellow region), according to Eqs. (\ref{eq:Pe}) and (\ref{eq:Pe2}). The frequency is normalized by $f_c=1.9 \rm Hz$, in the left panel, as given by Eq. (\ref{eq:e6}), and $f_c=7 \rm Hz$ in the right panel, as given by eq. (\ref{eq:ei3}). In both cases we assume an equal-mass sBH binary, with total mass $m=50M_\odot$, orbiting a SMBH, of mass $M=5\times10^6M_\odot$, at $\widetilde{a}=10^3$, in units of its Schwarzschild radius, or $0.5 \rm mpc$. For the dispersion-dominated case we assume $v/v_H=10$. The maximal frequency, $f_{ISCO}$, corresponds to a circular orbit at the binary's ISCO, at about $3R_s$. The axis above the figures presents the frequency in Hz, and the arrows show the frequency band of different GW detectors: LVK, ET and LISA.}
    \label{fig:6}
\end{figure*}

Given our fiducial parameters, as appears in Eq. (\ref{eq:e6}), the probability to retain eccentricity of $e\geq 0.5$ at an orbital frequency that corresponds to $f=10 \rm Hz$, the entrance to the LVK band, is about $30\%$.
Note that the assumed distance of the binary from the SMBH, which is stated in units of the SMBH's Schwarzschild radius, is equivalent to $\sim0.5 \rm mpc$. The probability decreases at larger orbital distances, e.g., for a binary at $0.5 \rm pc$ we get that the probability to retain $e\geq0.5$ is $10\%$.

\subsection{Gravitational-wave emission from eccentric orbits} \label{sec:strain}

As mentioned above, the GW emission from highly eccentric orbits is qualitatively different from the emission from circular ones.
The GWs from circular orbits are roughly periodic in the time-domain with a narrow peak in the frequency-domain\footnote{E.g., for $e\leq0.1$, more than $\sim90\%$ of the emitted power is radiated at twice the orbital frequency \citep[where we used the results derived by][]{Peters_1963}.}.

In contrast, GWs from highly eccentric orbits appear as a set of bursts in the time-domain, separated by the shrinking orbital period, with a broadband spectrum in the frequency-domain \citep{Peters_1963,Turner_1977}, covering all the harmonics of the orbital period and peaking at\footnote{For a more accurate estimation of the peak frequency see \cite{Hamers_2021}.}
\begin{equation}\label{eq:fp}
    f_p\sim\frac{1}{\pi}\sqrt{\frac{Gm}{r_p^3}}.
\end{equation}

All of the sBH binaries, formed via GW emission, follow an initially highly eccentric orbit and coalesce within the LVK observing time, as the merger time of the sBH binaries \citep{Peters_64} is of order of a month, given our fiducial parameters:
\begin{equation}\label{eq:Tmerger}
    \begin{aligned}
        T_{\rm Merger}\lesssim &\frac{192\sqrt{2}}{85}\kappa^{1/47}\frac{R_s}{c}\frac{m}{\mu}\left(\frac{r_{p,\rm GW}}{Rs}\right)^4\sqrt{\frac{R_H}{r_{p,\rm GW}}}\\
        & \sim0.5 \rm month \left(\frac{\widetilde{a}}{10^3}\right)^{63/47}\left(\frac{m}{50M_\odot}\right)\\
        &\quad \times\left(\frac{10^{-5}}{m/M}\right)^{42/47}\left(\frac{\mu/m}{1/4}\right)^{3/47},
    \end{aligned}
\end{equation}
where we used our estimate of the maximal periapsis, Eq. (\ref{eq:e3a}), and its corresponding semimajor axis, as follows from Eq. (\ref{eq:tg5}). Note that the merger time is always bound by the orbital period at the Hill radius, which is roughly the orbital period of the binary around the SMBH.

Considering the observed GW signal, we distinguish between binaries that circularize before entering the observed frequency band and those that form inside it. 
The latter group is defined by their initial peak frequency, as described in Eq. (\ref{eq:fp}), being higher than the minimal observed frequency, $f_{\rm obs,min}$, allowing for their identification as highly eccentric.
Therefore, the probability to measure highly eccentric orbits is given by
\begin{equation} \label{eq:Pecc}
    \begin{aligned}
        P_{\rm ecc}\simeq&\left(\frac{f_c}{f_{\rm obs, min}}\right)^{1/3}\approx 0.6\left(\frac{10 \rm Hz}{f_{\rm obs,min}}\right)^{1/3}\left(\frac{10^3}{\widetilde{a}}\right)^{6/47}\\
        & \times \left(\frac{50M_\odot}{m}\right)^{1/3}\left(\frac{m/M}{10^{-5}}\right)^{4/47}\left(\frac{1/4}{\mu/m}\right)^{7/47},
    \end{aligned}
\end{equation}
where we used Eqs. (\ref{eq:gw8}), (\ref{eq:e6}), and (\ref{eq:fp}).

Note that these orbits may result in a direct plunge if their periapsis is smaller than $2R_s$, the minimal periapsis of a bound orbit around a Schwarzschild BH.
Therefore, the probability for a direct plunge is given by
\begin{equation}\label{eq:e8a}
    \begin{aligned}
        P_{\rm plunge}\approx& 0.2\left(\frac{10^3}{\widetilde{a}}\right)^{6/47}\left(\frac{m/M}{10^{-5}}\right)^{4/47}\left(\frac{1/4}{\mu/m}\right)^{7/47},
    \end{aligned}
\end{equation}
as determined by Eq. (\ref{eq:gw8}) with $r_p=2R_s$. 

In conclusion, given our fiducial parameters, roughly $\sim40\%$ of the sBH binaries formed by GW emission in a shear-dominated disk circularize to smaller eccentricities ($e\lesssim0.3$) before entering the LVK band, while $\sim60\%$ form in the observed band and thus can be measured as highly eccentric. However, about third of the eccentric orbits lead to a direct plunge rather than an inspiral.

\section{Effects of initial eccentricity and inclination} \label{sec:e2}

In the above calculation we assume that the orbits of the sBHs are coplanar and therefore the motion is restricted to a $2D$ plane. As mentioned above, this assumption is motivated by the prediction that AGN disks tend to align the sBHs orbits \citep{Syer_1991,Generozov_2023}. 
However, qualitatively, there are two competing effects, dissipation induced by the disk and excitation due to encounters. Depending on the details of the gaseous disk and the distribution of the BHs, which are highly uncertain in the context of AGN, there can be two characteristic steady-state scenarios: shear-dominated and dispersion-dominated dynamics \citep[for further details see][]{Goldreich_04}. In the first case, dissipation is dominant over the excitation and thus the velocity dispersion $v$ is smaller than\footnote{This range corresponds to different distances from the SMBH, from $0.5 \rm pc$ down to $0.5 \rm mpc$.} $v_H\sim1-100\ \rm km/s$, the typical eccentricities are smaller than $e_H\sim R_H/a\sim 0.01$, and the inclinations are negligible. In this case, the relative velocities between the sBHs is determined by the Keplerian shear between circular orbits with different radii, and our above calculation is valid. 

In the second case, where $v\gtrsim v_H$, the excitation by close encounters has an important role and the typical relative velocities stem from the velocity dispersion rather than the shear in the disk. In this case, the eccentricities and inclinations are non-negligible and comparable, since strong scatterings can turn in-plane velocity (given by eccentricity) to a velocity perpendicular to the plane (given by inclination). Therefore, the dynamics are in $3D$ and close encounters can be treated as two-body interactions, neglecting the effects of the SMBH's tidal force because of the high relative velocities.

The $3D$ nature of the dynamics yields $p(\Delta b)\sim \Delta b$, as the band of impact parameters leading to capture forms a disk centered around the ``zero-angular momentum" impact parameter\footnote{This can also be understood from angular momentum conservation, as discussed in \cite{Li_Lai_Rodet_2022}.}, in contrast to $p(\Delta b)=const.$ in the $2D$ case. Analogously to the derivation of Eq. (\ref{eq:gw6}), conservation of angular momentum yields
\begin{equation}\label{eq:ei0}
    \frac{r_p}{R_H}\sim\left(\frac{v}{v_H}\frac{\Delta b}{R_H}\right)^2,
\end{equation}
and therefore
\begin{equation}\label{eq:ei1}
    P\left(<r_p\right)\propto\frac{r_p}{R_H},
\end{equation}
in agreement with \cite{Li_Lai_Rodet_2022}. 

Additionally, the critical periapsis distance for capture is smaller than in the shear-dominated case, Eq. (\ref{eq:gw5}), as now $\Delta E_{GW}\sim \mu v^2$ and so
\begin{equation}\label{eq:ei2}
    \frac{r_{p,\rm cap}}{R_H}\simeq \left(\frac{v}{v_H}\right)^{-4/7}\kappa^{2/7},
\end{equation}
in accordance with the results of \cite{OLeary_2009} and \cite{Samsing_2020}. The corresponding capture cross-section is given by
\begin{equation}\label{eq:ei2a}
    \frac{\sigma}{R_H^2}\sim\left(\frac{v}{v_H}\right)^{-18/7}\kappa^{2/7},
\end{equation}
and the characteristic frequency, analogues to Eq. (\ref{eq:e6}), is
\begin{equation}\label{eq:ei3}
    \begin{aligned}
        f_{c}\approx & 7{\rm Hz} \left(\frac{v/v_H}{10}\right)^{6/7}\left(\frac{10^3}{\widetilde{a}}\right)^{3/7} \\
        &\times \left(\frac{50M_\odot}{m}\right)\left(\frac{m/M}{10^{-5}}\right)^{2/7}\left(\frac{1/4}{\mu/m}\right)^{3/7}.
    \end{aligned}
\end{equation}

Thus, in the dispersion-dominated case, the minimal eccentricity, $e_{\rm min}$, is larger, since $r_{p,\rm cap}$ is smaller, yet the distribution is more strongly dominated by orbits with $r_{p,i}\sim r_{p,\rm cap}$, as evident by comparing Eqs. (\ref{eq:ei1}) and (\ref{eq:ei2}) with Eqs. (\ref{eq:gw5}) and (\ref{eq:gw8}).

The eccentricity cdf is given by
\begin{equation}\label{eq:Pe2}
    \begin{aligned}
        P(\geq e)\simeq &\left(\frac{f_c}{f}\right)^{2/3}\frac{1}{g(e)}\\
        &\times\frac{g\left(e\right)-g\left(e_{\rm max}\right)}{g\left(e_{\rm min}\right)-g\left(e_{\rm max}\right)},
    \end{aligned}
\end{equation}
where $e_{\rm min}$ is given by Eq. (\ref{eq:emin}), and $e_{\rm max}$ is given by
\begin{equation} \label{eq:emaxv}
    \begin{aligned}
        \left(1-e_{\rm max}\right)\sim&\kappa^{4/49}\left(\frac{f}{f_c}\right)^{10/21}\left(\frac{v}{v_H}\right)^{20/49}\\
        &\times \left[1+\left(\frac{f_c}{f}\right)^{2/3}\kappa^{2/7}\left(\frac{v}{v_H}\right)^{10/7}\right]^{-2/7},
    \end{aligned}
\end{equation}
which is the dispersion-dominated equivalent of Eq. (\ref{eq:emax}).

For $e\ll1$, we get from Eqs. (\ref{eq:Pe2}) and (\ref{eq:e3}) that 
\begin{equation}
    \begin{aligned}   
         P(\geq e)\simeq& \left(\frac{f_c}{f}\right)^{2/3}\frac{1}{g\left(e\right)}\\
        \approx&\frac{0.2}{e^{12/19}}\left(\frac{v/v_H}{10}\right)^{4/7}\left(\frac{50\rm Hz}{f}\right)^{2/3}\left(\frac{10^3}{\widetilde{a}}\right)^{2/7}\\
        & \times \left(\frac{50M_\odot}{m}\right)^{2/3}\left(\frac{m/M}{10^{-5}}\right)^{4/21}\left(\frac{1/4}{\mu/m}\right)^{2/7},
    \end{aligned}
\end{equation}
while for $e\simeq1$
\begin{equation}\label{eq:ei4}
    \begin{aligned}
        P(\gtrsim e\simeq1)&\simeq\left(\frac{f_c}{f}\right)^{2/3}\left(e_{\rm max}-e\right).
    \end{aligned}
\end{equation}

In the right panel of Fig. \ref{fig:6} we summarize the probabilities for circular and eccentric orbits as a function of the orbital frequency in the dispersion-dominated regime. Considering the LVK band, the probability to retain eccentricity of $e\geq0.5$ at $10 \rm Hz$ is about $40\%$, for sBH binaries at $0.5 \rm mpc$, and $5\%$ for binaries at $0.5\rm pc$. 

Following the discussion in section \ref{sec:strain}, we get the dispersion-dominated equivalents of Eqs. (\ref{eq:Pecc}) and (\ref{eq:e8a}): the probability to measure an eccentric merger at a given observed frequency band is
\begin{equation}\label{eq:Peccw}
    \begin{aligned}
        P_{\rm ecc}\simeq&\left(\frac{f_c}{f_{\rm obs, min}}\right)^{2/3}\approx 0.8\left(\frac{v/v_H}{10}\right)^{4/7}\left(\frac{10 \rm Hz}{f_{\rm obs,min}}\right)^{2/3}\\
        & \times \left(\frac{10^3}{\widetilde{a}}\right)^{2/7}\left(\frac{50M_\odot}{m}\right)^{2/3}\left(\frac{m/M}{10^{-5}}\right)^{4/21}\left(\frac{1/4}{\mu/m}\right)^{2/7},
    \end{aligned}
\end{equation}
and the probability for a direct plunge is
\begin{equation}\label{eq:ei5b}
    \begin{aligned}
        P_{\rm plunge}\approx& 0.1\left(\frac{v/v_H}{10}\right)^{4/7}\left(\frac{10^3}{\widetilde{a}}\right)^{2/7}\\
        & \times\left(\frac{m/M}{10^{-5}}\right)^{4/21}\left(\frac{1/4}{\mu/m}\right)^{2/7}.
    \end{aligned}
\end{equation}

Thus, in the dispersion-dominated case there is a higher probability to measure eccentric mergers in the LVK band, compared to the shear-dominated case, with smaller fraction of direct plunges. Nonetheless, in this case there is a stronger dependence on the binary's orbital radius and therefore the probability significantly decreases when considering binaries at larger distances from the SMBH.
\section{Summary \& Discussion} \label{sec:sum}

In this work, we study analytically the effects of tidal forces (from the SMBH) and GW emission on sBH binary captures in AGN disks. We estimate the capture cross-section and present an order of magnitude study of the three-body dynamics, involving two sBHs and an SMBH, in the shear-dominated regime, which is verified by numerical integration of the Hill equations with an effective GW-induced friction force.

We further study the post-capture eccentricity evolution. We identify a critical observing frequency, $f_c$, below which only eccentric mergers will be measured (as shown in the lower panel of figure \ref{fig:5}), and above which low eccentricity mergers will be detected, with a small probability tail extending to high eccentricities (upper panel of figure \ref{fig:5}). 
For sBHs with coplanar and circular initial orbits, i.e., in a shear-dominated disk, the critical frequency, given our fiducial parameters, is about $2\rm Hz$ (see Eq. \ref{eq:e6}),
and $\sim40\%$ of the binaries circularize before entering the LVK band, while the rest can be measured as highly eccentric, detectable from their formation, with $e\approx1$, up to the merger. However, A considerable fraction of the eccentric mergers, $\sim 1/3$, leads to a direct plunge rather than an eccentric inspiral.

For eccentric and inclined initial orbits, where the disk is dispersion-dominated, the critical frequency is larger, $7 \rm Hz$, for our fiducial parameters (see Eq. \ref{eq:ei3}). Consequently, the probability for measuring eccentric orbits at the LVK band is larger, $\sim80\%$, given our fiducial parameters.

We note that in order to predict the total detection rate of GW sources from this formation channel, the distribution and characteristics of AGN should be taken into account. In addition, in this work we focus on binaries formed by GW capture, without considering the effects of the surrounding gas in AGN disks, which modifies the sBHs orbital evolution and hence the eccentricity distribution and the total formation rate.

Future earth-based GW detectors, such as the Einstein Telescope \citep[ET;][]{ET_2020}, will be sensitive to lower frequencies, from $\sim1\rm Hz$. Therefore, ET will detect a larger fraction of the sBH binary mergers before they circularize. 
Furthermore, the space-based GW detector LISA \citep{LISA} will be sensitive to both the low-frequency tail of these highly eccentric mergers and wider eccentric binaries. Combining measurements from LISA with those from ground-based detectors as ET and LVK will allow a detailed, multiband study of the sBHs binaries formation mechanisms.

\begin{acknowledgments}
This research was partially supported by an ISF grant, an NSF/BSF grant, and an MOS grant. B.R. acknowledges support from the Milner Foundation. 
\end{acknowledgments}

\bibliography{main}{}
\bibliographystyle{aasjournal}

\end{document}